# Deep Learning Interfacial Momentum Closures in Coarse-Mesh CFD Two-Phase Flow Simulation Using Validation Data

Han Bao[1], Jinyong Feng[2,*], Nam Dinh[3], Hongbin Zhang[1]

[1]Idaho National Laboratory, P.O.Box 1625, MS 3860, Idaho Falls, 83415, Idaho, USA

[2]Massachusetts Institute of Technology, Cambridge, 02139, Massachusetts, USA

[3]North Carolina State University, Raleigh, 27695, North Carolina, USA

## Abstract

Multiphase flow phenomena have been widely observed in the industrial applications, yet it remains a challenging unsolved problem. Three-dimensional computational fluid dynamics (CFD) approaches resolve of the flow fields on finer spatial and temporal scales, which can complement dedicated experimental study. However, closures must be introduced to reflect the underlying physics in multiphase flow. Among them, the interfacial forces, including drag, lift, turbulent-dispersion and wall-lubrication forces, play an important role in bubble distribution and migration in liquid-vapor two-phase flows. Development of those closures traditionally rely on the experimental data and analytical derivation with simplified assumptions that usually cannot deliver a universal solution across a wide range of flow conditions. In this paper, a data-driven approach, named as feature-similarity measurement (FSM), is developed and applied to improve the simulation capability of two-phase flow with coarse-mesh CFD approach. Interfacial momentum transfer in adiabatic bubbly flow serves as the focus of the present study. Both a mature and a simplified set of interfacial closures are taken as the low-fidelity data. Validation data (including relevant experimental data and validated fine-mesh CFD simulations results) are adopted as high-fidelity data. Qualitative and quantitative analysis are performed in this paper. These reveal that FSM can substantially improve the prediction of the coarse-mesh CFD model, regardless of the choice of interfacial closures, and it provides scalability and consistency across discontinuous flow regimes. It demonstrates that data-driven methods can aid the multiphase flow modeling by exploring the connections between local physical features and simulation errors.

---

[*] Corresponding author. Tel,: +1 919 645 8225

*E-mail address*: fengjinyong2008@gmail.com

**Acronyms**

| | | | |
|---|---|---|---|
| BAMF | bubbly and moderate void fraction | IT | interface tracking |
| CFD | computational fluid dynamics | KDE | kernel density estimation |
| DFNN | deep feedforward neural network | LF | low-fidelity |
| DNB | departure from nucleate boiling | ML | machine learning |
| DNS | direct numerical simulation | NPP | nuclear power plant |
| FSM | feature similarity measurement | NRMSE | normalized root mean squared error |
| | | PWR | pressurized water reactor |
| GELI | global extrapolation through local interpolation | QoI | quantity of interest |
| HF | high-fidelity | RANS | Reynolds-averaged Navier–Stokes |
| IC/BC | initial condition/boundary condition | | |

## 1. Introduction

As a carbon-free energy source, nuclear power plants (NPPs) play an important role in the reliable supply of energy, national security, and environmental impact. However, since the 1970s, the number of NPPs under construction in the United States has gradually dropped due to high capital and construction costs and increased safety standards. It is imperative to leverage advanced numerical models and simulations, which can provide advanced suggestions for reactor design and complement dedicated experimental testing, to reduce the design and construction cost of NPPs. Recently, three-dimensional thermal hydraulics methods, in the form of computational fluid dynamics (CFD), have made tremendous advancements. They promise to transform the way in which we approach the design of more efficient and reliable systems. One of the most challenging, yet widely encountered phenomena extant in NPPs is multiphase flow—i.e., liquid and vapor in complex phase interactions of mass, momentum, and energy.

Various modeling approaches have been introduced and applied on multiphase flow based on specific research goals and on-hand computational costs. Among them, on a spectrum toward first-principle methods is direct numerical simulation (DNS) coupled with interface tracking (IT) (Hirt and Nichols, 1981; Sussman et al., 1994; Unverdi and Tryggvason, 1992), where bubble-liquid interfaces are resolved on a sufficiently fine scale. While the DNS/IT methods can be leveraged to study separate effects on single bubble (Bunner and Tryggvason, 2003; Feng and Bolotnov, 2017a, 2017b, 2017c), practical two-phase flow simulations with on the order of hundreds of bubbles are still not computationally affordable (Fang et al., 2018). Compared to the DNS/IT methods, the most general framework for multiphase flow is the Eulerian-Eulerian two-fluid model, which combines Reynolds-averaged Navier-Stokes (RANS)-based models to model turbulence with a variety of interfacial closure laws to account for the exchange of mass, momentum, and energy between individual phases. The Eulerian–Eulerian two-fluid approach assumes that all phases coexist inside each computational cell. For each fluid, the full set of conservation equations is solved; therefore, each fluid has a different velocity field.

Based on ensemble-averaging techniques, the Eulerian-Eulerian two-fluid model can operate on a relatively coarse mesh and is comparatively computationally efficient for industrial applications. As an example of the computational costs involved, $5 \times 5$ pressurized water reactor (PWR) fuel bundle with mixing vanes requires ~170 million cells to generate consistent mesh resolution and obtain reasonable results for the prediction of departure from nucleate boiling (DNB) (Brewster et al., 2019). It is important to explore the options of adopting coarse-mesh CFD methods for the practical industrial applications.

Using coarse-mesh CFD methods is expected to have two major sources of error: physical-model error and mesh-induced error. Physical-model error arises from physical assumptions and mathematical approximations in the form of closure models and errors in assigning values for model parameters or calibration coefficients. For adiabatic two-phase flow, one of the most-significant approximations is the formulation of interfacial forces. Drag force originates from the balance between buoyancy and gravitational forces. Despite the difficulty in modeling these physical complexities, two phase modeling is not possible wihtout interfacial drag closure. Lateral distribution of the bubble is a combination effect of lift, turbulent-dispersion, and wall-lubrication forces where those three forces are model treatments based on relevant physical observations. Although driven by dedicated experiments (Dijkhuizen et al., 2010; Tomiyama et al., 2002) and analytical derivation (Antal et al., 1991; Lubchenko et al., 2018), the expression of those forces may not be consistent and universal across a wide range of flow conditions. Physical experiments have high construction cost for full-size prototype facilities and scaling issues in using scaled-down facilities. These costs limit the validation and uncertainty quantification of the CFD models. While new techniques are continuously developed and deployed, the availability of an experimental database is still limited by the state-of-the-art measurement techniques. For example, measurements of bubbles size and distribution are not available for reactor operating pressures.

Distinct from physical-model errors, mesh-induced error comes from the solution discretization in space and time and the approximation used in over-cell integration of variables that are non-uniformly distributed in the cell. Ideally, adopting finer mesh would approach the realistic representations of physical phenomena, but the computational cost requires attention. However, when the mesh size is much smaller than the bubble size, the ensemble-averaging assumption for bubbly flow will be no longer hold. Physical-model and mesh-induced error are tightly connected. For example, the calculation of flow variable gradients depends on the mesh resolution between two adjacent cells, thus inaccuracies in gradients from truncation error will augment the physical model errors of a RANS turbulence closure model.

Recently, machine-learning (ML)-based methods have emerged as a valuable approach to aid the development and application of CFD methods. Different ML algorithms, such as neural networks and random forests, have been widely applied to predict relevant parameters or source terms in turbulence-transport equations (Milani et al., 2019; Tracey et al., 2015; Xiao et al., 2020). These ML applications mainly focus on the improvement of RANS turbulence modeling for the single phase, without considering mesh-induced numerical uncertainties and resultant errors. Turbulence-model error is quantified in different validation domains, but the scalability of these data-driven models needs further demonstration. In this paper, scalability is defined as the predictive capability of a model or code for extrapolative conditions, such as boundary condition, geometry or flow regime. Another pioneering work that utilizes local physical features from high-fidelity (HF) DNS simulation to predict boiling heat transfer is conducted by Liu et al. (Liu et al., 2018); therein, mesh-induced numerical error is assumed to be similarly small. Aiming at correcting mesh-induced numerical error without considering the model errors, a coarse-gird CFD approach was proposed by Hanna et al. (Hanna et al., 2020) for the prediction of local simulation errors. These existing efforts estimate model error and mesh-induced numerical error separately to achieve a better predictive performance; however, they ignore the tight connection between these two main simulation-error sources, as described in the last paragraph. For two-phase flow using CFD or other system codes, local mesh size is treated as one of the key model parameters, and fine-mesh convergence is somehow not expected. The uncertainty propagation resulting from scaling distortions makes it more difficult to estimate and reduce simulation errors for modeling and simulation of realistic system-level NPP analysis.

In this paper, a recently proposed data-driven approach, feature-similarity measurement (FSM), takes model error, mesh-induced numerical error, and scaling distortions into consideration by treating the physical correlations, coarse mesh sizes, and numerical solvers as an integrated model. This FSM approach uses deep learning to explore the relationship between specific physical feature groups and simulation variables. The well-trained deep-learning model can be considered as a surrogate for governing equations and closure correlations of coarse-mesh CFD. Coarse-mesh cases with both a mature set of interfacial force closures, i.e., bubbly and moderate void fraction (BAMF) (Sugrue et al., 2017), and a simplified set of interfacial-force closures are trained with the neutral network with limited experimental data and HF CFD models. The research presented here paves the path for resolving the challenges of modeling interfacial-force closures with the aid of FSM.

## 2. Methodology

Feature Similarity Measurement (FSM), developed by Bao et al. (Bao et al., 2020, 2019c, 2019a, 2019b, 2018a), integrates model error, mesh-induced error and scaling uncertainty together, and estimates the simulation error by exploring local patterns in multiscale data with the use of deep learning. The underlying local patterns in multi-scale data are represented by a set of physical features, defined based on physical systems of interest, empirical correlations, and local mesh size. After generating an error database using validation data (including relevant experimental data and validated fine-mesh CFD simulations results) and low-fidelity (LF) simulation data (fast-running coarse-mesh simulation results), deep-learning algorithms are applied to explore the relationship between the local physical features and local simulation errors to develop a surrogate model. For the applications of the FSM approach, a turbulent-mixing case study was performed to evaluate the predictive capability of FSM for globally extrapolated conditions in (Bao et al., 2018a). In (Bao et al., 2020, 2019c), FSM was applied for a two-dimensional mixed convection problem considering turbulence. Case studies show that FSM has good predictive capability, and the prediction accuracy increases with increasing data similarity of physical features. A data-driven framework was developed to improve applications of the coarse-mesh codes by predicting their simulation errors and suggesting the optimal mesh size and closure models for system-level simulations based on the FSM approach (Bao et al., 2019a). In (Bao et al., 2019b) FSM was improved as data-driven guidance to improve the coarse-mesh CFD simulation capability for two-phase flow with a substantially reduced computational cost. Currently FSM has been applied for different thermal-hydraulic codes such as GOTHIC and STAR-CCM+. GOTHIC is a system thermal-hydraulic simulation tool that is developed and widely used for containment analysis (Bao et al., 2018b; Chen and Yuann, 2015; EPRI, 2014).

### 2.1. Summary of feature similarity measurement

The basic idea of FSM was proposed in (Bao et al., 2018a) as shown in Figure 1. FSM includes three steps: establishing local similarity by identifying local physical features, measuring the similarity of local physical features, and enhancing local similarity to improve the prediction performance of ML models. The key hypothesis of FSM is that the well-trained deep feedforward neural network (DFNN) model has better predictive performance if (1) the size of training data is sufficient for the DFNN model to capture underlying local patterns; (2) the DFNN model is sufficiently complicated to learn and capture the local patterns in the training data; and (3) the similarity between training data and testing data is high enough for the DFNN model to provide accurate predictions based on local patterns learned from the training data. For different target systems (i.e., phenomena, structure/geometry, and initial/boundary conditions [IC/BCs]), the requirements on the levels of data sufficiency, ML-model uncertainty, and data similarity are different. The quantification of the acceptable criteria and evaluation metrics will be further studied. Meanwhile, a concept called global extrapolation through local interpolation (GELI) was proposed in (Bao et al., 2020) to indicate the situation where the global physical condition of a target case is identified as an extrapolation of existing cases the local physics of which are similar. The extrapolation of global physics indicates different boundary conditions, different geometries, structures, or dimensions, or other different global physical conditions, such as a set of characteristic non-dimensional parameters. The interpolation of local physics implies that the underlying local physics of existing cases is assumed to be represented by a set of physical features, and the physical-feature data of the target case are mostly covered or similar to the physical feature data of existing cases.

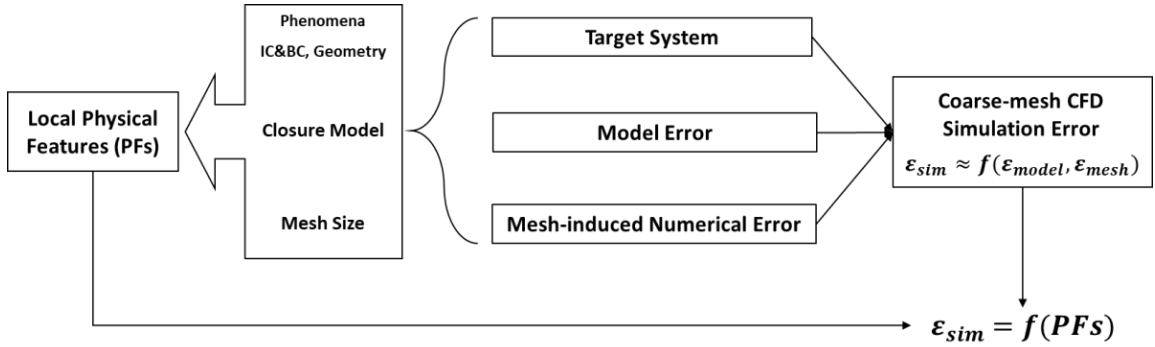

Figure 1. Basic idea of FSM: identify the relationship between physical features and simulation error.

To take regional information into consideration, the physical-feature group should include the gradients of local variables and local physical parameters that are able to represent local physical behaviors or to be applied in crucial closure relationships, as shown in Figure 2. More details about how to identify physical features can be found in (Bao et al., 2019a, 2019b).

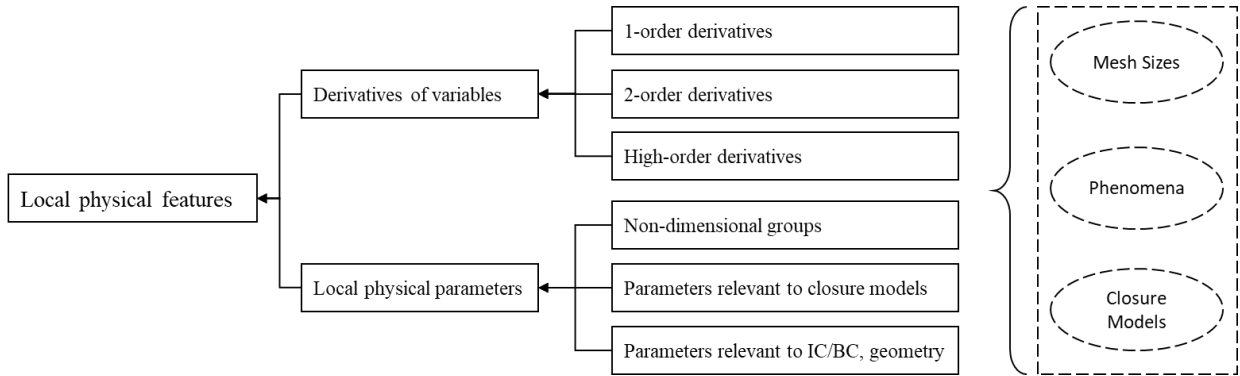

Figure 2. Classification of local physical features.

After identifying local physical features and generating low- and high-fidelity data, a DFNN model can be trained to define the function of local physical features and local simulation errors of quantities of interest (QoIs). The FSM workflow for the training, prediction, and validation of the DFNN model is illustrated in Figure 3. To investigate the predictive performance of FSM in two-phase flow, data are divided into two parts for training and validation. The training dataset consists only of physical features $PF_i$ that are calculated based on LF data and local simulation error $\varepsilon_i$ that are defined as the discrepancy between HF data $QoI_{HF,i}$ and LF data $QoI_{LF,i}$. Then deep ML techniques are applied for the training of a data-driven surrogate model. In this work, DFNN is used as the ML tool, but other ML algorithms can also be used, depending on the nonlinearity of the target problem. Once obtaining the DFNN model $\varepsilon = f(PF)$, predictions can be made by inserting new values of physical features in validation data $PF_j$ as $\boldsymbol{\varepsilon_j} = f(\boldsymbol{PF_j})$. Then DFNN-predicted data can be calculated using $\boldsymbol{QoI_{pre,j}} = \boldsymbol{QoI_{LF,j}} + \boldsymbol{\varepsilon_j}$. The error of DFNN prediction can be calculated using the metric as normalized root mean squared errors (NRMSE), as expressed in Equation (1).

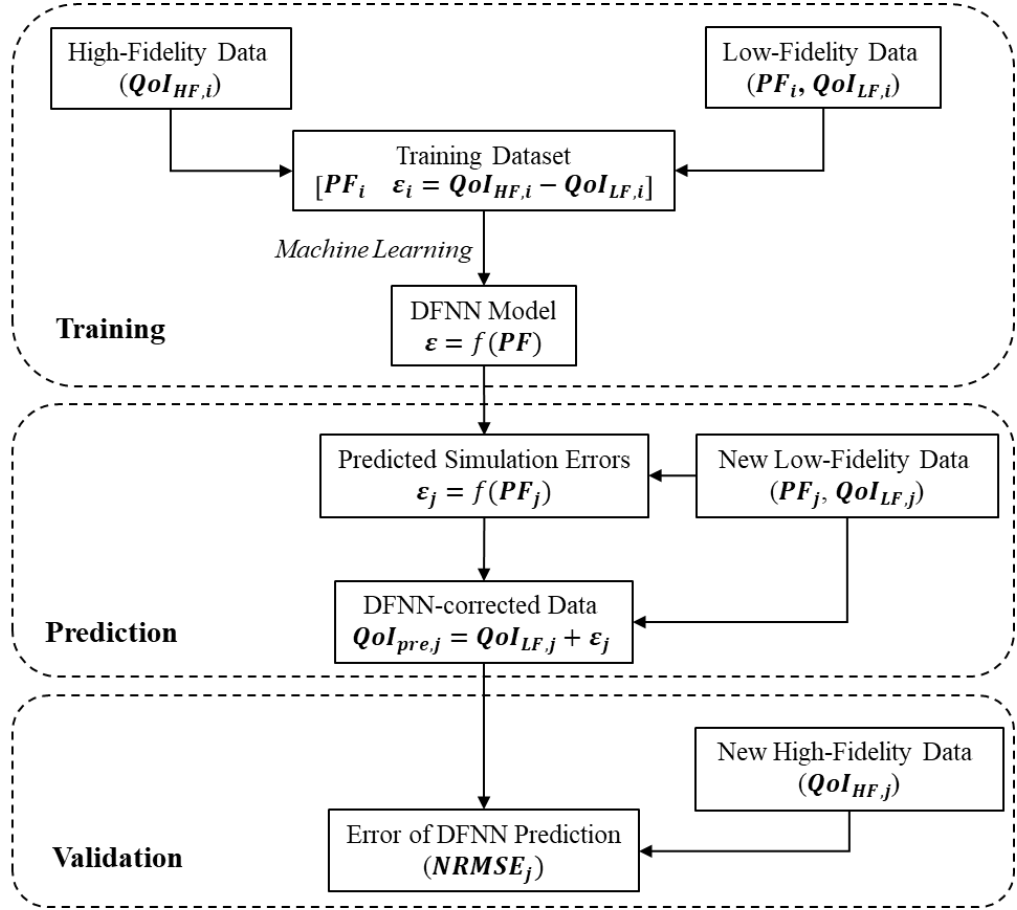

Figure 3. Training, prediction and validation of DFNN model in FSM application.

$$NRMSE^k = \frac{\sqrt{\frac{1}{n}\sum\left(QoI_{HF}^k - QoI_{pre}^k\right)^2}}{\frac{1}{n}\sum QoI_{HF}^k} \qquad (1)$$

where $QoIs$ in this two-phase case are two velocities ($u, v$) and temperature ($T$), $k = 1, 2, 3$ respectively represents liquid and vapor velocities ($u_l$ and $u_g$), and void fraction ($\alpha$). By using the DFNN model, the FSM approach can predict the simulation errors in LF predictions and obtain the DFNN-corrected predictions $QoI_{pre}^k$ and respective prediction errors $NRMSE^k$.

Data similarity between training data and target data is measured using a method called kernel-density estimation (KDE), which is a non-parametric way to estimate the probability-density function. KDE assumes the training-data distribution can be approximated as a sum of multivariate Gaussians. One can use kernel distribution when a parametric distribution cannot properly describe the data or when one wants to avoid making unnecessary assumptions about the distribution of data. KDE can be considered as the probability that the point ($q$) locates in the distribution of training data ($p_{i,j}, i = 1,2, \dots, n; j = 1,2, \dots, d$). It is expressed as (Scott, 2015),

$$p_{KDE} = \frac{1}{n \cdot h_1 h_2 \dots h_d} \sum_{i=1}^{n} \prod_{j=1}^{d} k(\frac{q_j - p_{i,j}}{h_j}) \qquad (2)$$

where $d$ is the number of variables in $q$ and $p_i$, k is the kernel-smoothing function, and $h_j$ is the bandwidth for each variable. A multivariate kernel distribution is defined by a smoothing function ($k$) and a bandwidth matrix defined by $H = h_1, h_2, \dots, h_d$, which controls the smoothness of the resulting density curve. Therefore, KDE can be used to measure the distance by estimating the probability of a given point located in a set of training-data points. Then the similarity between training data ($p_{i,j}, i = 1,2, \dots, n; j = 1,2, \dots, d$) and target data ($q_{j,k}, j = 1,2, \dots, d; k = 1,2, \dots, m$) is expressed as the mean of KDEs $S_{KDE}$. A greater value of $S_{KDE}$ means a higher level of similarity.

$$S_{KDE} = \frac{1}{m} \sum_{k=1}^{m} p_{KDE,k} = \sum_{k=1}^{m} \frac{1}{n \cdot h_1 h_2 \dots h_d} \sum_{i=1}^{n} \prod_{j=1}^{d} ker(\frac{q_{j,k} - p_{i,j}}{h_j}) \qquad (3)$$

The process for enhancing local data similarity between training data and target data will be discussed in the Section 3.

## 2.2. Deep feedforward neural network

This work uses a DFNN to fit the function between simulation error and several physical-feature inputs. A DFNN normally includes several hidden layers with a couple of neurons and activation functions on each hidden layer. By using multiple layers of transformations, deep neural networks are able to capture complex, hierarchical interactions between features. A typical three-hidden-layer feedforward network with one input layer, three hidden layers, and one output layer is shown in Figure 4. The network shown above has $\boldsymbol{p}_{R*1}$ inputs, $S_1$ neurons in the first layer, $S_2$ and $S_3$ neurons in the second and third layer, etc. Each layer (layer $i$) has a weight matrix $\boldsymbol{W}_{S_i*S_{i-1}}$, a bias vector $\boldsymbol{b}_{S_i*1}$, and an output vector $\boldsymbol{a}_{S_i*1}$. $\boldsymbol{W}_{S_i*S_{i-1}}$ and $\boldsymbol{b}_{S_i*1}$ are both adjustable scalar parameters of the neuron. The output of the third layer ($\boldsymbol{a}^3$) is the network output of interest, labeled as $\boldsymbol{y}$, which is a function of inputs $\boldsymbol{p}$ fitted by a DFNN model. The three-hidden-layer DFNN includes the following non-linear transformations:

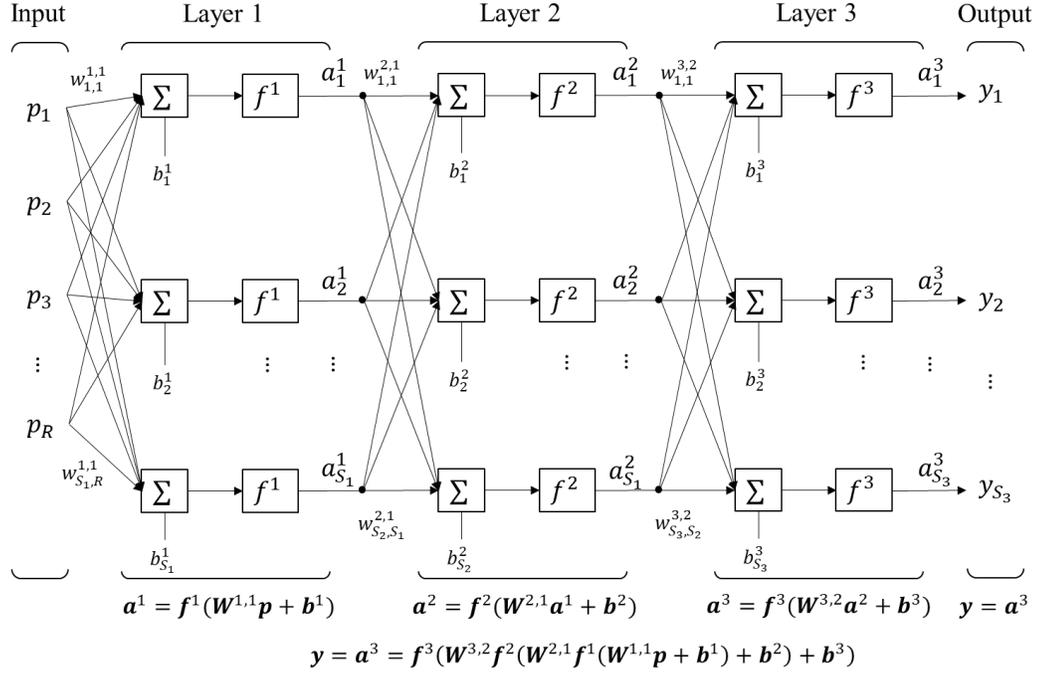

Figure 4. A typical three-hidden-layer feedforward network.

$$\boldsymbol{a}^1 = \boldsymbol{f}^1(\boldsymbol{W}^{1,1}\boldsymbol{p} + \boldsymbol{b}^1) \tag{4}$$

$$\boldsymbol{a}^2 = \boldsymbol{f}^2(\boldsymbol{W}^{2,1}\boldsymbol{a}^1 + \boldsymbol{b}^2) \tag{5}$$

$$\boldsymbol{a}^3 = \boldsymbol{f}^3(\boldsymbol{W}^{3,2}\boldsymbol{a}^2 + \boldsymbol{b}^3) \tag{6}$$

$$\boldsymbol{y} = \boldsymbol{a}^3 = \boldsymbol{f}^3(\boldsymbol{W}^{3,2}\boldsymbol{f}^2(\boldsymbol{W}^{2,1}\boldsymbol{f}^1(\boldsymbol{W}^{1,1}\boldsymbol{p} + \boldsymbol{b}^1) + \boldsymbol{b}^2) + \boldsymbol{b}^3) \tag{7}$$

Here $\boldsymbol{f}^i$ is an activation function that produces the output $\boldsymbol{a}^i$. The tanh-sigmoid activation function is commonly used in backpropagation networks because it is differentiable and non-linear. Once the network weights and biases are initialized, the network is ready for training. During training, the weights and biases of the network are iteratively adjusted to minimize the network prediction error. The evaluation metric of the function fitting is the mean squared error at each test point.

$$mse = \frac{1}{N_d}\sum_{k=1}^{N_d}(\boldsymbol{t}_k - \boldsymbol{y}_k)^2 \tag{8}$$

where $\boldsymbol{t}_k$ and $\boldsymbol{y}_k$ represent real data and respective prediction. The gradient of cost function (error) is applied to determine how to adjust the weights for minimizing the error. The gradient is determined using a technique called backpropagation, which involves error computations backwards through the network by generalizing the gradient descendent rule to multiple-layer networks and nonlinear, differentiable activation functions. In this work, Bayesian regularization (MacKay, 1992) is applied as the training function to adjust weights and biases. This algorithm prevents overfitting on the training set, a problem which if not avoided, prevents the DFNN from being well generalized to new data outside the training set. It obtains a better solution for some noisy and challenging problems. More details about Bayesian regularization algorithm for backward propagation derived by MacKay (MacKay, 1992) and Burden (Burden and Winkler, 2008) can be found in references.

## 3. Formulation of case study

In all previous applications of FSM, HF data were generated using CFD simulations having validated interfacial closures coupled with fine-mesh configuration. It is difficult to combine specific appropriate empirical closures and local mesh sizes to validate these "high-fidelity" CFD simulations, particularly when the experimental data has large ranges of global conditions. These conditions could include largely different injection velocities and void fraction for a pipe flow, which indicates the existence of various flow regimes. No single set of closures has the capability to accurately predict all different conditions. Therefore, this paper proposes to include experimental data as well as HF CFD simulations in a HF database. A DFNN was trained using training datasets to predict the simulation errors of coarse-mesh LF simulations with HF data. By adding the predicted simulation errors to the original coarse-mesh LF simulation results, the DFNN-corrected results can be obtained and compared with HF data. The goal of this case study is to answer two questions: (1) Is FSM able to improve the coarse-mesh LF simulations to reach the accuracy of HF data? (2) Does FSM work well for different types of LF and HF data?

In this paper, a case study based on two-phase flow was performed to evaluate the predictive capability of FSM. Four tests were designed using different HF and LF data. Forty-two reference experimental datasets are from Liu and Bankoff (Liu and Bankoff, 1993a, 1993b). Inflow conditions are summarized in Table *1*. Two different types of validation data, HF-EXP and HF-CFD, are used as HF data in this case study. Experimental data are used as HF-EXP data. HF-CFD data were generated using a commercial CFD package, STAR-CCM+13.02 ("STAR-CCM+, version 13.02," 2018), with 25-cell mesh configuration and the following two-phase interfacial forces closures and turbulence models: a drag-force model from Tomiyama (Tomiyama et al., 1998), lift correction from Shaver and Podowski (Shaver and Podowski, 2015) with a base coefficient of 0.025, turbulent-dispersion force from Burns (Burns et al., 2004) and the standard $k - \varepsilon$ turbulence model. This set of closures, which is referred as the BAMF model, has been tested for 12 cases from the Liu and Bankoff experimental datasets, provides reasonable predictions for mean flow profiles of void fraction and phase velocities (Sugrue et al., 2017). In two-fluid model, interfacial momentum exchange terms are introduced to represent the interaction between bubbles and surrounding liquid. Driven by the physical observations, such as bubble migration in shear flow and flat void fraction profile caused by highly turbulent flow, turbulent dispersion, lift and other forces are derived and introduced to represent the underlying physics. While those interfacial forces have their own model form error, advanced experiment and interface resolving numerical simulations help the development of interfacial forces and make those closure approach the ideal solution.

To accelerate the simulation, only one-quarter of the domain is simulated, and symmetric boundary conditions are applied on the two side surfaces. The cross-sectional view of the mesh configurations is shown in Figure 5. QoIs in this case study are liquid and vapor velocities ($u_l$ and $u_g$), and void fraction ($\alpha$).

Table 1. Summary of the flow conditions of experimental cases

| Set | Injection Superficial Velocity | | Void fraction $\alpha$ | Set | Injection Superficial Velocity | | Void fraction $\alpha$ |
|---|---|---|---|---|---|---|---|
| | $j_l$ (m/s) | $j_g$ (m/s) | | | $j_l$ (m/s) | $j_g$ (m/s) | |
| 1 | 0.376 | 0.027 | 0.0407 | 22 | 0.974 | 0.027 | 0.0204 |
| 2 | 0.376 | 0.067 | 0.1167 | 23 | 0.974 | 0.067 | 0.0514 |
| 3 | 0.376 | 0.112 | 0.1843 | 24 | 0.974 | 0.112 | 0.0791 |
| 4 | 0.376 | 0.18 | 0.2449 | 25 | 0.974 | 0.18 | 0.1242 |
| 5 | 0.376 | 0.23 | 0.3079 | 26 | 0.974 | 0.23 | 0.1512 |
| 6 | 0.376 | 0.293 | 0.3657 | 27 | 0.974 | 0.293 | 0.1869 |
| 7 | 0.376 | 0.347 | 0.4168 | 28 | 0.974 | 0.347 | 0.2108 |
| 8 | 0.535 | 0.027 | 0.0312 | 29 | 1.087 | 0.027 | 0.0176 |
| 9 | 0.535 | 0.067 | 0.0877 | 30 | 1.087 | 0.067 | 0.0473 |
| 10 | 0.535 | 0.112 | 0.1406 | 31 | 1.087 | 0.112 | 0.0737 |
| 11 | 0.535 | 0.18 | 0.2016 | 32 | 1.087 | 0.18 | 0.1096 |
| 12 | 0.535 | 0.23 | 0.2344 | 33 | 1.087 | 0.23 | 0.1497 |
| 13 | 0.535 | 0.293 | 0.3102 | 34 | 1.087 | 0.293 | 0.1777 |
| 14 | 0.535 | 0.347 | 0.3398 | 35 | 1.087 | 0.347 | 0.1976 |
| 15 | 0.753 | 0.027 | 0.0235 | 36 | 1.391 | 0.027 | 0.0148 |
| 16 | 0.753 | 0.067 | 0.0622 | 37 | 1.391 | 0.067 | 0.0387 |
| 17 | 0.753 | 0.112 | 0.1091 | 38 | 1.391 | 0.112 | 0.0581 |
| 18 | 0.753 | 0.18 | 0.1554 | 39 | 1.391 | 0.18 | 0.0964 |
| 19 | 0.753 | 0.23 | 0.1816 | 40 | 1.391 | 0.23 | 0.1176 |
| 20 | 0.753 | 0.293 | 0.2381 | 41 | 1.391 | 0.293 | 0.1504 |
| 21 | 0.753 | 0.347 | 0.2692 | 42 | 1.391 | 0.347 | 0.1724 |

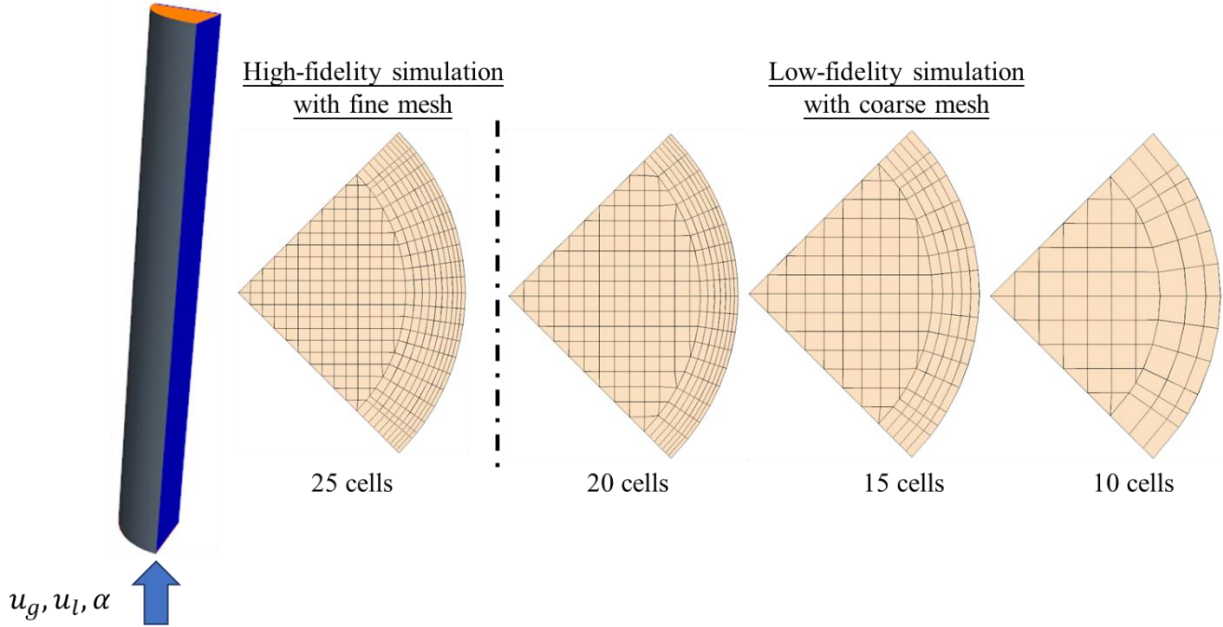

Figure 5. Mesh configuration for HF and LF simulations using STAR-CCM+.

LF data were generated using STAR-CCM+ with three different coarse meshes, as shown in Figure 5. The number of cells from the wall to the pipe center ranges from 10 to 20. Mesh configuration is crucial to the accurate prediction of the multiphase-flow phenomena using CFD approaches. The calculation of flow-variable gradients depends on the mesh resolution between two adjacent cells, which are directly relevant to certain physical models. For example, lift force relies on the gradient of velocity and the magnitude of the turbulence-dispersion force, which depends on the gradient of void fraction. In addition, the near-wall mesh resolution determines numerous closures near the wall, like velocity wall function. The lift-coefficient model proposed by Shaver and Podowski (Shaver and Podowski, 2015) also depends on the ratio between wall distance and interaction length scale.

LF-Type-I data are generated by "zero" interfacial closures, which only solves basic momentum equations and a standard two-equation $k - \varepsilon$ turbulence model without any interfacial closures except Tomiyama's drag coefficient (Tomiyama et al., 1998), data are generated without any model treatment and only have the naturally emerged drag force. LF-Type-II data are generated using a mature set of interfacial-force models that are referred to as the BAMF model. Closures in the BAMF model were selected based on numerical stability, simplicity, cross-code portability, and robustness, which provide a mature representation of the bubble migration behavior in bubbly flow. In this way, the FSM method can virtually drive the bubble migration within the Eulerian-Eulerian two-fluid model and see whether ML algorithms can improve coarse mesh CFD predictions and supplement non-linear complex, yet extremely important, interfacial momentum closures. A summary of the interfacial closures used in LF-Type-I and II data generation are given in Table 2.

Table 2. Comparison of the low fidelity Type I and Type II interfacial momentum closures.

| Model | | LF-Type-I | LF-Type-II and HF-CFD |
|---|---|---|---|
| Turbulence model | | Standard $k - \varepsilon$ linear | |
| Interfacial momentum forces | Drag coefficient | Tomiyama (Tomiyama et al., 1998) | |
| | Lift coefficient | N/A | Shaver and Podowski (Shaver and Podowski, 2015) |
| | Turbulent dispersion force | N/A | Burns (Burns et al., 2004) |
| | Wall lubrication force | N/A | Shaver and Podowski's correction (Shaver and Podowski, 2015) |

In summary, four tests are conducted to demonstrate the prediction capability of FSM approaches where the HF data have two separate groups of experimental data and fine-mesh CFD simulation. Coarse-mesh LF data are composed with a mature and a simplified set of interfacial closures. All 42 cases have different values for void fraction and phase velocities. Two cases are selected separately as a testing case. Case 7 with the lowest liquid rate and highest vapor rate and void fraction is used as Test Case 1, Case 42 with the highest injection velocities and highest void fraction is selected as Testing Case 2. For each test case, the other 41 cases will be used to generate training data for correcting the 10-cell simulation results of testing cases. Therefore, for each test, there are 1845 ($41 \times (25+20+15)$) data points for training and 10 data points for testing. Four tests are formalized in this paper using different high-fidelity datasets and LF closures, as listed in Table 3. Considering LF and HF data are generated using different approaches (simulation or experiment), or different closures and mesh configurations, a large discrepancy between LF- and HF data is expected.

Table 3. Tests with different HF and LF data types.

| Data Type (LF: low fidelity, HF: high fidelity) | LF-Type-I (Coarse-mesh Simplified closures) | LF-Type-II (Coarse-mesh BAMF) |
|---|---|---|
| HF-CFD (Fine-mesh BAMF) | Test 1 | Test 2 |
| HF-EXP (Experimental data) | Test 3 | Test 4 |

## 4. Results

### 4.1. Local similarity establishment

#### 4.1.1. Identification of local physical features

For this case study, 27 total physical features are defined according to the involved phenomena, IC/BCs, structure/geometry, and the closure models applied in data generation, as listed in Table 4. Sixteen of these physical features are the 1-order and 2-order derivatives of key variables, including liquid and vapor velocity ($u_l$ and $u_g$), void fraction ($\alpha$), pressure ($P$), liquid and vapor kinetic energy ($k_l$ and $k_g$), liquid and vapor turbulence dissipation rate ($\varepsilon_l$ and $\varepsilon_g$). Another 11 physical features are local physical parameters defined based on the model parameters and mesh sizes. As many relevant parameters as possible are included, and the importance (weight and bias) of each parameter is justified by the ML algorithms. The non-dimensional group contains six classic non-dimensional parameters. The Reynolds number is defined as the ratio between the inertial forces to viscous forces. In this work, three different local Reynolds numbers, $Re_\Delta$, $Re_b$, and $Re_y$, are defined respectively using local mesh size ($\Delta$), pre-set bubble size ($D_b$) and wall distance ($y$) as characteristic lengths, which take the effects of mesh size, closure model, and boundary condition into consideration. Turbulent intensity ($I_l$ and $I_g$) provides a measurement of the flow fluctuations versus the mean flow velocity. The Weber number ($We$) characterizes the relative importance of the fluid's inertia compared to its surface tension. The parameters relevant to closure models, IC/BCs, and geometry include $R_l$ and $R_g$ because the ratio between turbulent length scale, i.e., $\frac{k^{\frac{3}{2}}}{\varepsilon}$, and the bubble diameter, $D_b$; non-dimensional wall distance, $R_b$, considering the velocity and void-fraction distributions, crucially depend on the wall distance; $r_l$ representing the ratio of liquid turbulent eddy viscosity and liquid molecular viscosity, which is supposed to become important for the high Reynolds number regime; and $R_\mu$ as the ratio between the gas- and liquid-eddy viscosity, which characterizes the magnitude of modeled turbulence level for liquid and gas.

Table 4. Identification of physical feature for bubbly flow CFD simulation.

| Derivatives of variable | | | | Local physical parameters | | | |
|---|---|---|---|---|---|---|---|
| 1-order derivatives | | 2-order derivatives | | Non-dimensional groups | | Parameters relevant to closure models, IC/BC, geometry | |
| $\dfrac{du_l}{dx}$ | $\dfrac{du_g}{dx}$ | $\dfrac{d^2u_l}{dx^2}$ | $\dfrac{d^2u_g}{dx^2}$ | $Re_\Delta = \dfrac{\rho_l \Delta \cdot \Delta u}{\mu_l}$ | $I_l = \dfrac{k_l}{u_l^2}$ | $R_l = \dfrac{k_l^{\frac{3}{2}}}{\varepsilon_l D_b}$ | $R_\mu = \dfrac{\mu_g^t}{\mu_L^t}$ |
| $\dfrac{d\alpha}{dx}$ | $\dfrac{dP}{dx}$ | $\dfrac{d^2\alpha}{dx^2}$ | $\dfrac{d^2P}{dx^2}$ | $Re_b = \dfrac{\rho_l D_b \Delta u}{\mu_l}$ | $I_g = \dfrac{k_g}{u_g^2}$ | $R_g = \dfrac{k_g^{\frac{3}{2}}}{\varepsilon_g D_b}$ | $r_l = \dfrac{\mu_l^t}{\mu_l}$ |
| $\dfrac{dk_l}{dx}$ | $\dfrac{dk_g}{dx}$ | $\dfrac{d^2k_l}{dx^2}$ | $\dfrac{d^2k_g}{dx^2}$ | $Re_y = \dfrac{\rho_l y \Delta u}{\mu_l}$ | $We = \dfrac{\rho D_b \Delta u^2}{\sigma}$ | $R_b = \dfrac{D_b}{\Delta}$ | |
| $\dfrac{d\varepsilon_l}{dx}$ | $\dfrac{d\varepsilon_g}{dx}$ | $\dfrac{d^2\varepsilon_l}{dx^2}$ | $\dfrac{d^2\varepsilon_g}{dx^2}$ | | | | |

*4.1.2. Training and prediction of deep machine learning model*

Once physical features are defined and calculated based on LF data, a point-to-point method is applied to calculate the simulation errors of local QoIs ($u_l$, $u_g$, $\alpha$) for two-phase flow. With 27 inputs and three outputs, a DFNN containing three hidden layers and 20 neurons in each hidden layer (i.e., 20-20-20) is trained and used to predict the 10-cell coarse-mesh simulation errors of testing case. Test 1 has fine-mesh BAMF prediction as HF data and coarse-mesh LF-Type-I closure prediction as LF data, Test 2 has experimental data as HF data and coarse-mesh LF-Type-I closure prediction as LF data. Test 3 has fine-mesh and coarse-mesh BAMF predictions as HF and LF data, Test 4 has experimental data as HF data and coarse-mesh BAMF prediction as LF data. Different experimental data (Exp), fine-mesh and coarse-mesh simulation data (25-cell HF Sim, 10-cell LF Sim) and ML predictions (10-cell ML) are compared.

Figure 6 and Figure 7 show the results of Tests 1 and 2 for Case 7. Both tests use LF-Type-I closure to generate LF data. The original 10-cell coarse-mesh LF results show different patterns with the HF data, especially for void fraction. For Test 1, higher void fraction occurs near the wall due to the positive lift coefficient adopted in the BAMF model. As expected, a flat void fraction profile is predicted using LF-Type-I closure because no interfacial closures are applied for the transverse direction. For Test 2, LF-Type-I closure has surprisingly better predictions than HF-CFD results. One reason is that the void fraction profile is flat for Case 7, which means that the interfacial forces on the transverse direction are balanced by each other. LF-Type-I does not have any interfacial closures on transverse direction which accidently reaches the optimum balance of the mutual interactions between bubbles and surrounding liquid and makes them outperform the BAMF model. With the aid of ML, accuracy of LF-Type-I simulation results is further improved. It shows that no matter which type of HF or LF data are used, FSM has the capability to capture their differences and correct LF data to the selected HF data. Effects of model-induced error and physical-model error are considered simultaneously, and the LF-simulation errors are reduced. After the training, the DFNN model well captures this flat pattern and provides an appropriate correction to match the HF results. The capability of capturing regional patterns results from identifying 1-order and 2-order derivatives of QoIs as the physical features because, not only the characteristic at this point, but also the connections of this point with its neighboring points are captured. LF simulations were greatly improved after FSM-aided corrections.

Figure 7 displays the results of Tests 3 and 4 for Case 7, both tests use LF-Type-II closure to generate LF data. Similar to Test 1 and 2, the accuracy of original 10-cell simulation results is significantly improved after being corrected by the predicted simulation errors from well-trained DFNN model. The original 10-cell LF simulation presents different patterns with unphysical "peaks" locating at the seventh, eighth, and ninth point, which are attributed to the overpredicted high void fraction on the near-wall cell. The overprediction of high void fraction also leads to the unexpected fluctuations in velocity profiles. The well-trained DFNN model correctly predicts the pattern of these unexpected "peaks" and provides an appropriate correction to match HF results in different types. The results indicate that these local physical features can represent local physics and provide sufficiently accurate prediction on simulation error of QoIs. Although the predictions for void fraction need further improvements, FSM represents good scalability on estimating the local simulation errors even for the extrapolation of global conditions.

Figure 8 and Figure 9 compare the results of Tests 1 through 4 for Case 42. The void-fraction profile of coarse-mesh LF simulation using LF-Type-I closure is still flat because there is no force in the transverse direction. However, it becomes much closer to the HF data with the correction of the DFNN model. No matter which type of LF or HF data are used, FSM has the ability to learn and reproduce the patterns for new conditions.

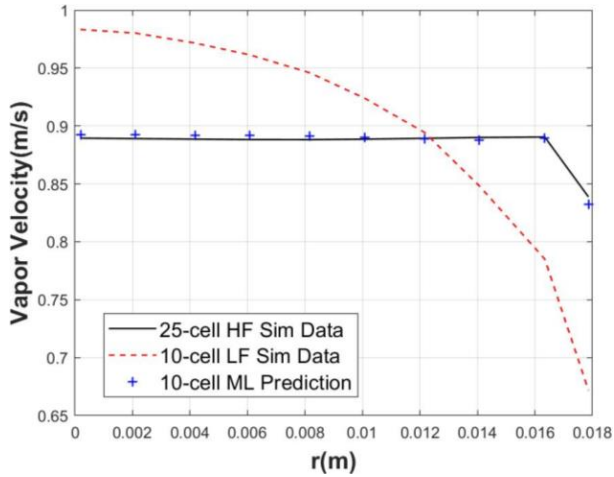

(a) Test 1: Vapor Velocity

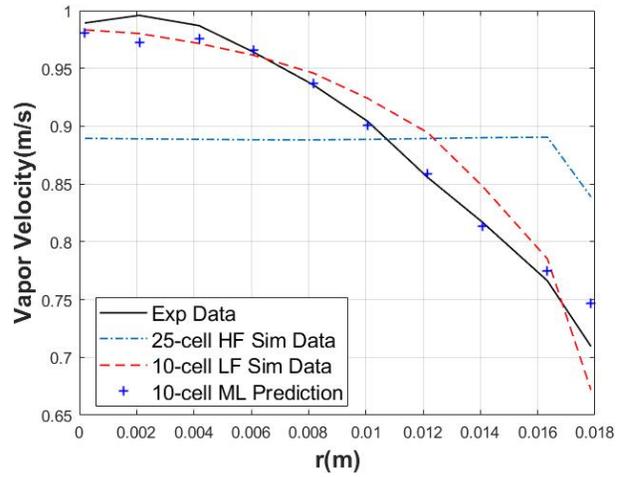

(d) Test 2: Vapor Velocity

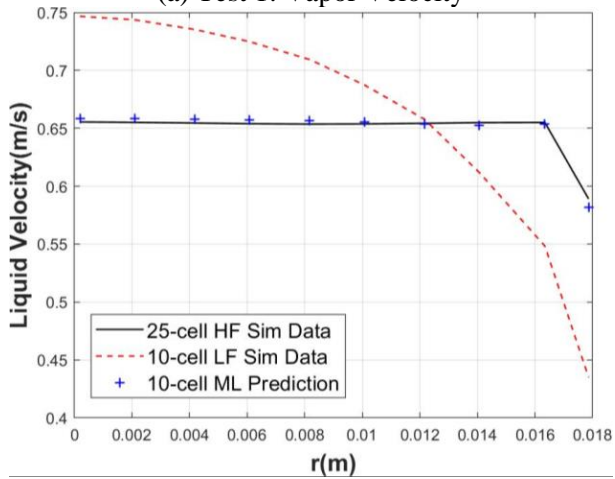

(b) Test 1: Liquid Velocity

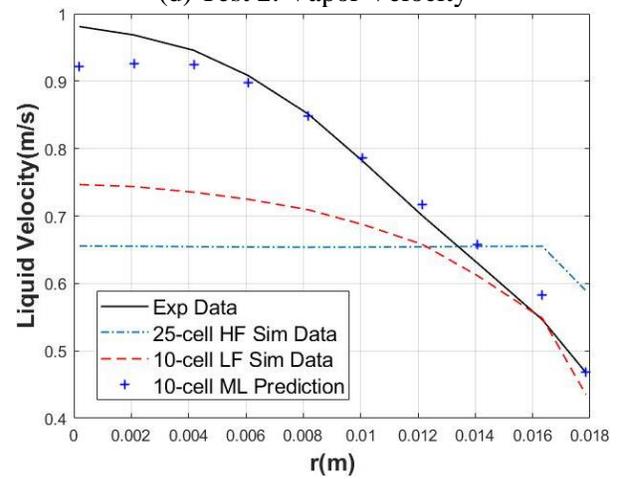

(e) Test 2: Liquid Velocity

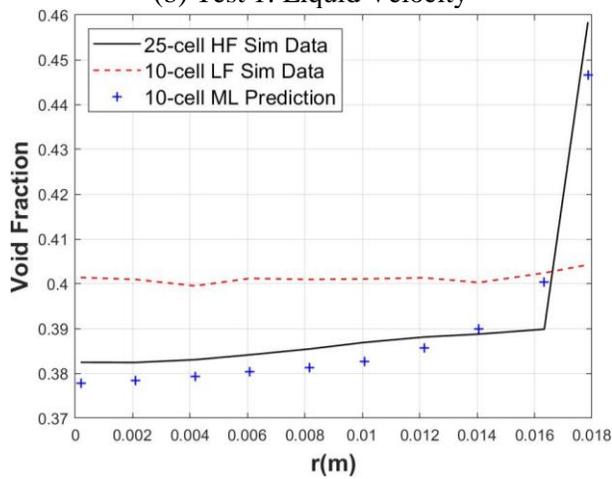

(c) Test 1: Void Fraction

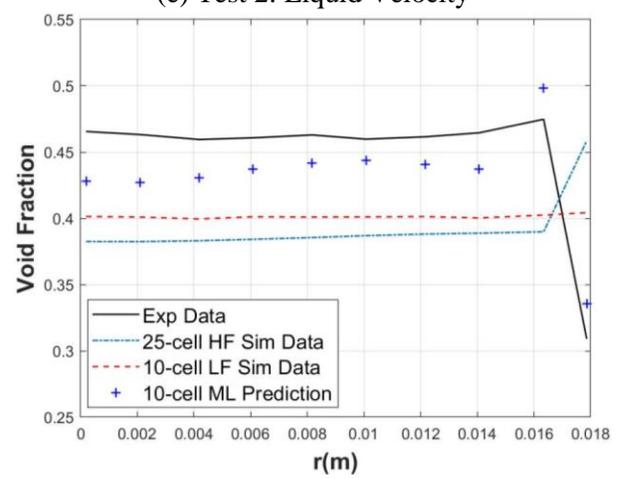

(f) Test 2: Void Fraction

Figure 6. Results of Test 1 and 2 for Case 7.

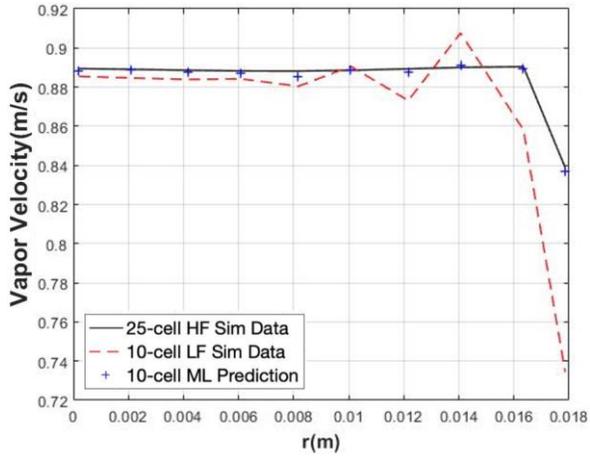

(a) Test 3: Vapor Velocity

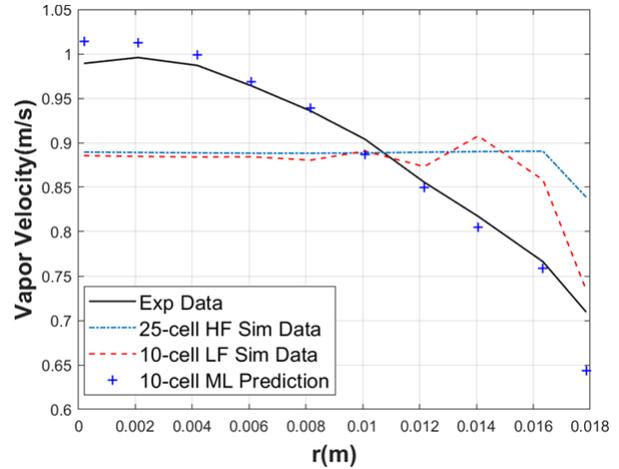

(d) Test 4: Vapor Velocity

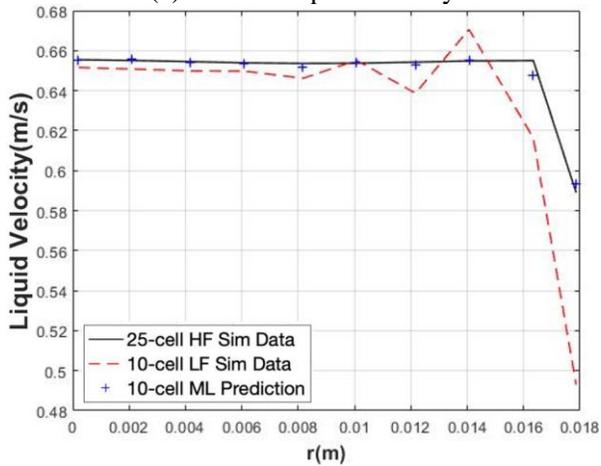

(b) Test 3: Liquid Velocity

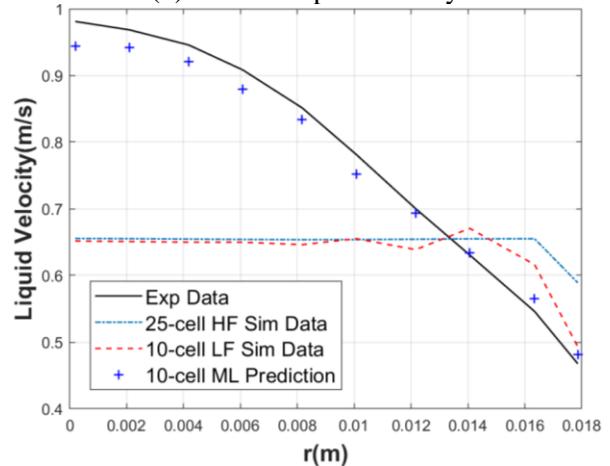

(e) Test 4: Liquid Velocity

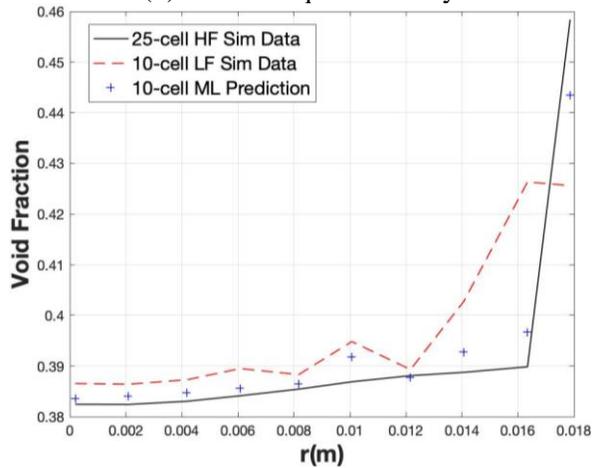

(c) Test 3: Void Fraction

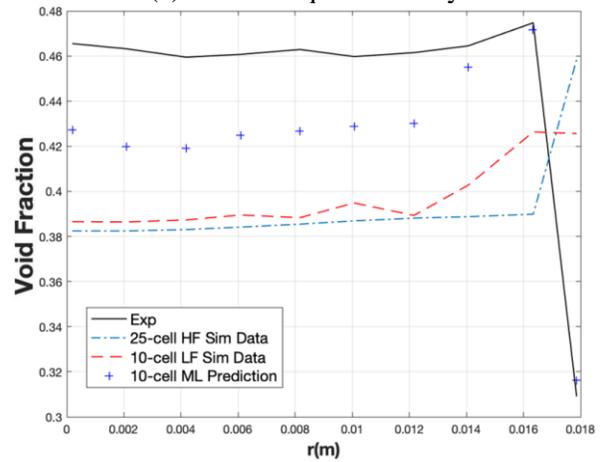

(f) Test 4: Void Fraction

Figure 7. Results of Test 3 and 4 for Case 7.

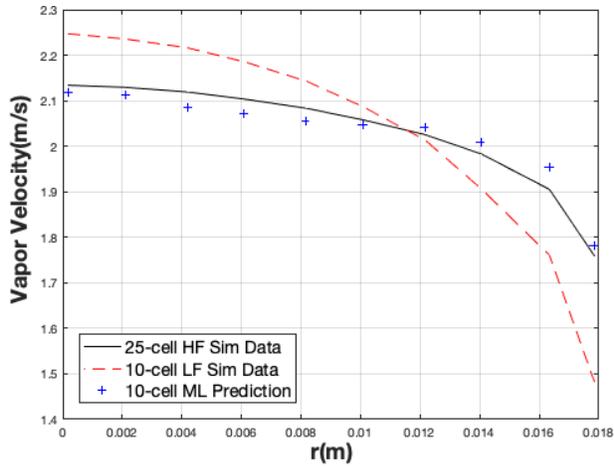

(a) Test 1: Vapor Velocity

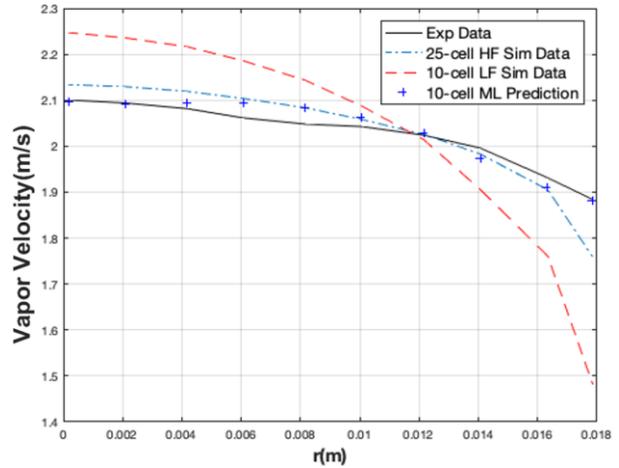

(d) Test 2: Vapor Velocity

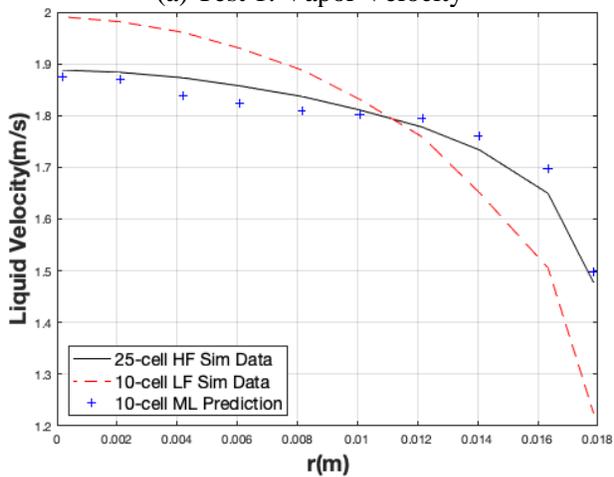

(b) Test 1: Liquid Velocity

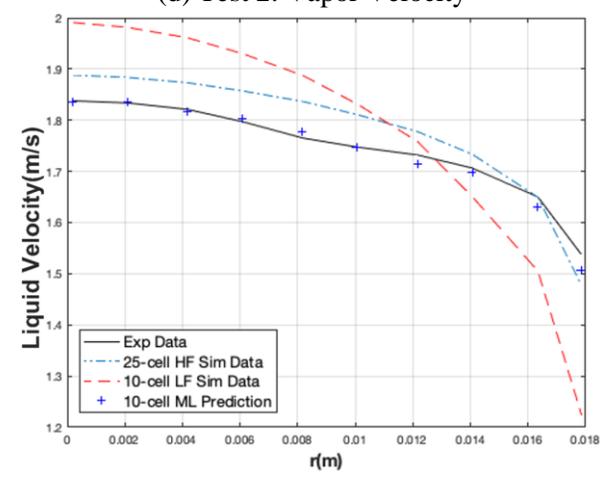

(e) Test 2: Liquid Velocity

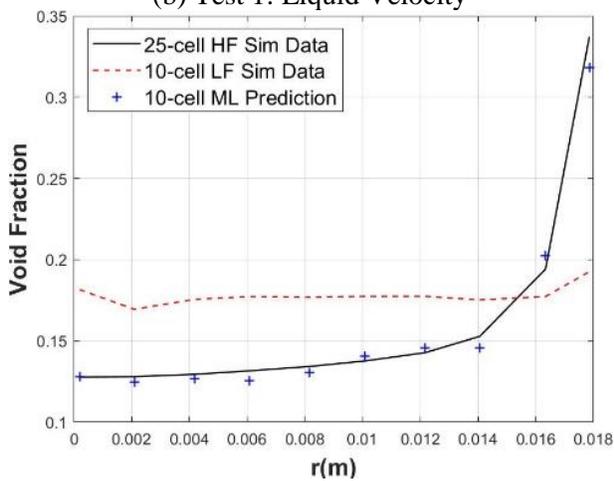

(c) Test 1: Void Fraction

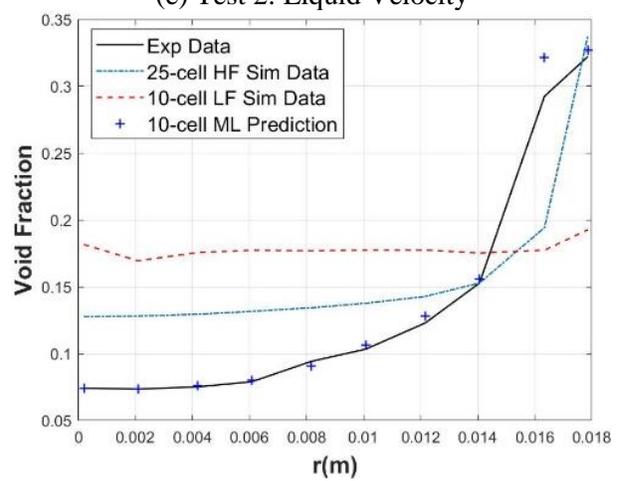

(f) Test 2: Void Fraction

Figure 8. Results of Test 1 and 2 for Case 42.

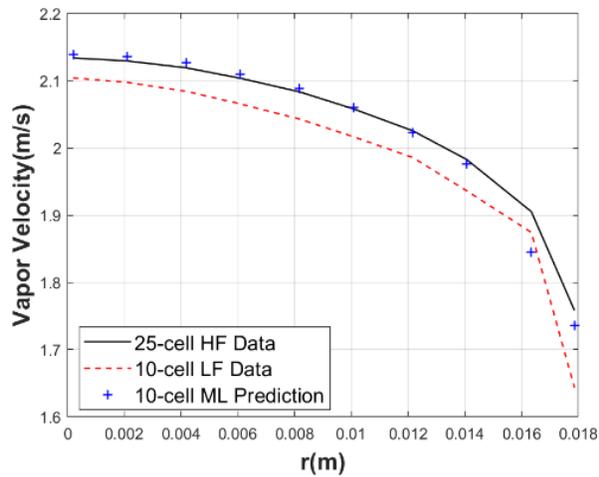

(a) Test 3: Vapor Velocity

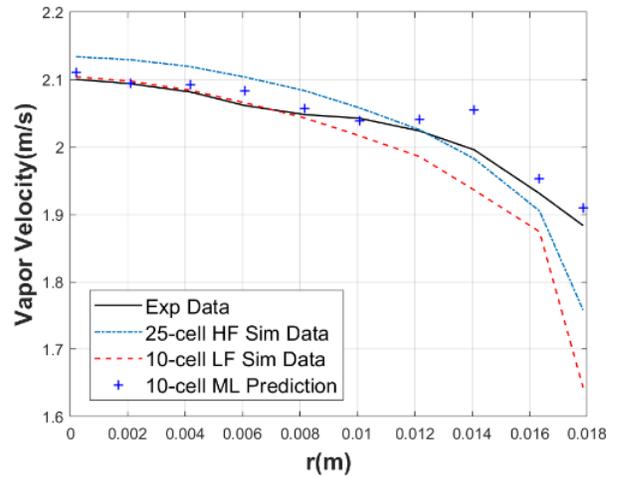

(d) Test 4: Vapor Velocity

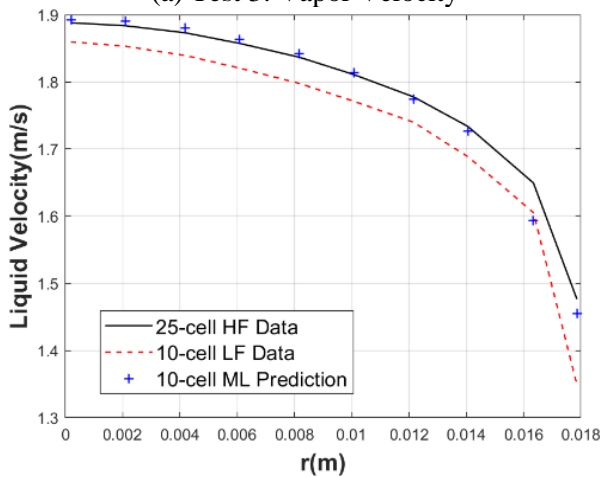

(b) Test 3: Liquid Velocity

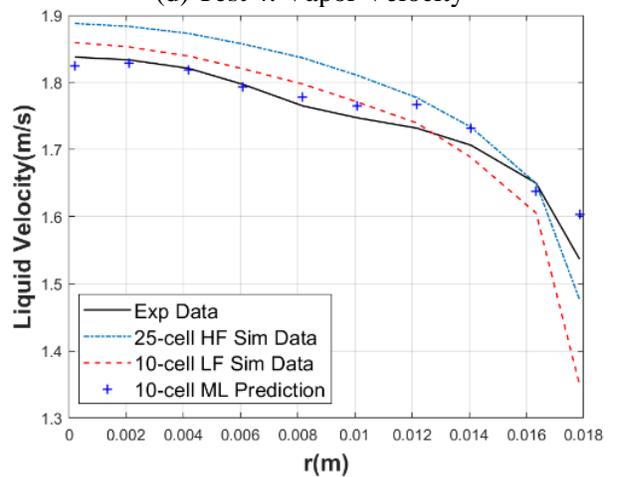

(e) Test 4: Liquid Velocity

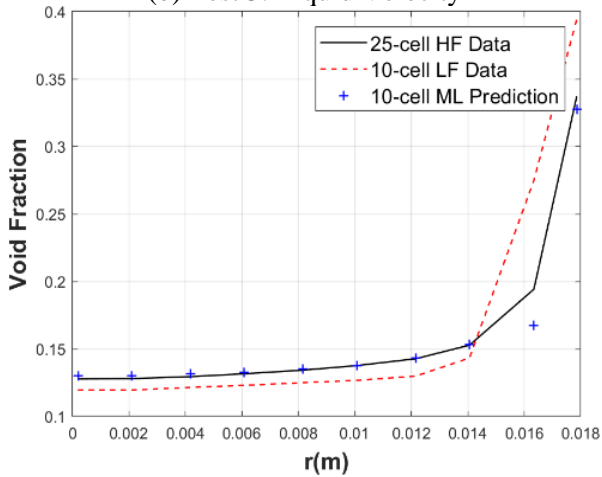

(c) Test 3: Void Fraction

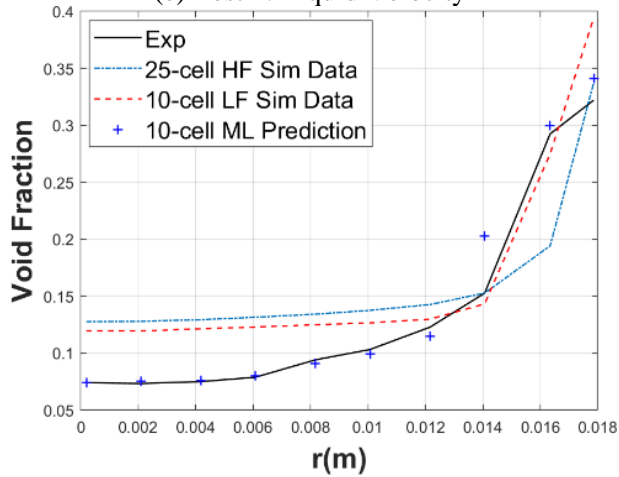

(f) Test 4: Void Fraction

Figure 9. Results of Test 3 and 4 for Case 42.

## 4.2. Local similarity measurement and enhancement

In Figure 6 (f), the DFNN-predicted void fraction for Case 7 is much closer to experimental data compared with the original LF prediction; however, it still needs further improvement. This section will illustrate an improvement process by measuring and enhancing local similarity between target data (10-cell simulation for Case 7) and training data (simulation for other 41 cases). As mentioned before, one of the key hypotheses of FSM is that the predictive performance of FSM will be improved with the increase of data similarity between enough training and testing data. Therefore, several "most similar" data points from original training dataset will be selected to construct an "optimal" training dataset with a smaller size, but higher similarity. In this case, by measuring the data distances between each data point in testing case (10-cell configuration) using Euclidean distance, $P$ data points with small values of data distance are selected for each target data point. There are totally $100P$ (or less if there are some points repeatedly counted) data points selected to build a new training dataset. After standardizing values of physical features into [-1, 1], Euclidean distance between single data points is calculated as expressed as below,

$$D_{m,q} = d(PF_m, PF_q) = \sqrt{\sum_{k=1}^{N}(x_{tr,k} - x_{ta,k})^2} \tag{9}$$

Provided that the number ($Q = 10$) of target data points is much smaller than the number ($M = 41 * 45 = 1845$) of data points in the data warehouse, $D_{m,q}$ is defined as the Euclidean distance between training data point $PF_m$ and target data point $PF_q$, $(1 \leq m \leq M \ and \ 1 \leq q \leq Q)$. $N$ is the number of physical features, $x_{tr,k}$ and $x_{ta,k}$ are respectively the values of physical feature number $k$ of $PF_m$ and $PF_q$. Therefore, the optimal training dataset that just includes these target-similar data points can be constructed. Table 5 summarizes the distribution of these target-similar data points in different mesh configurations with different values of $P$. With fewer selected training data points (lower $P$, from all to 100), the training dataset is built with higher data similarity ($S_{KDE}$) and the trained DFNN model has relatively better prediction (lower NRMSEs). However, when $P$ is too low (50), the number of selected data points is too small to satisfy the requirements of DFNN training. It meets the hypotheses of FSM that a balance between data similarity and training-data quantity should be achieved to obtain the "optimal" predictive capability of the DFNN model. For this two-phase flow case study, the optimal point may be reached when the value of $P$ is about 100. Void fraction predictions of Test 2 for Case 7 with different training datasets are shown in Figure 10, all the original LF predictions (red dashed lines) are corrected by adding DFNN-predicted simulation errors and approach to HF experimental data (black lines). Results in Figure 10 (a) and (b) are generated using different sizes of training dataset as $P = 100$ and $P = 200$, while all simulation data points of the other 41 cases are used for Figure 10 (c).

According to Table 5, when $P$ has a relatively low value (smaller than 600), the 10-cell and 15-cell data points have a higher level of similarity than the 20-cell data points because nearly all selected training-data points come from 10-cell and 15-cell simulations, even though there are more data points generated in 20-cell simulations. The value of mesh sizes significantly affects the calculations of local physical features. When $P = 800$, the number of 20-cell data points is higher than the number of 10-cell data points, but only about 80% (1493/1845) of-cell data points are selected while 89.5% (367/410) 10-cell data points are selected. It also implies that some data points from fine-mesh simulation results have lower similarity than some 10-cell data points with the target 10-cell data points. When only 15-cell and 20-cell simulation data points are used for building the training dataset, its similarity to the target 10-cell data is only 0.2147. However, the respective void-fraction prediction with the correction of DFNN predicted simulation error has a better performance than the original LF predictions, as shown in Table 5 and Figure 10 (d). The scale gap between fine meshes and coarse mesh is filled in FSM because the mesh effect is considered and captured by using mesh sizes for the calculation of local physical features.

Table 5. Distribution of target-similar data points in different mesh configurations.

| Target Case: Case 7 10-cell mesh configuration | | | | | | |
|---|---|---|---|---|---|---|
| # of target-similar data points for each target data point (P) | # of training data points from other 41 cases (no repeat) | | | | Data similarity ($S_{KDE}$) | NRMSE of prediction |
| | 10-cell | 15-cell | 20-cell | Total | | |
| 50 | 143 | 75 | 8 | 226 | 0.4228 | 0.0578 |
| 100 | 196 | 165 | 20 | 385 | 0.3867 | 0.0409 |
| 200 | 247 | 290 | 107 | 664 | 0.3638 | 0.0493 |
| 400 | 314 | 431 | 293 | 1038 | 0.3255 | 0.0485 |
| 600 | 331 | 551 | 473 | 1315 | 0.2980 | 0.0539 |
| 800 | 367 | 569 | 557 | 1493 | 0.2820 | 0.0553 |
| All | 410 | 615 | 820 | 1845 | 0.2390 | 0.0600 |
| Only 15-cell and 20-cell | 0 | 615 | 820 | 1435 | 0.2147 | 0.0622 |
| Original LF simulation | NA | | | | NA | 0.1487 |

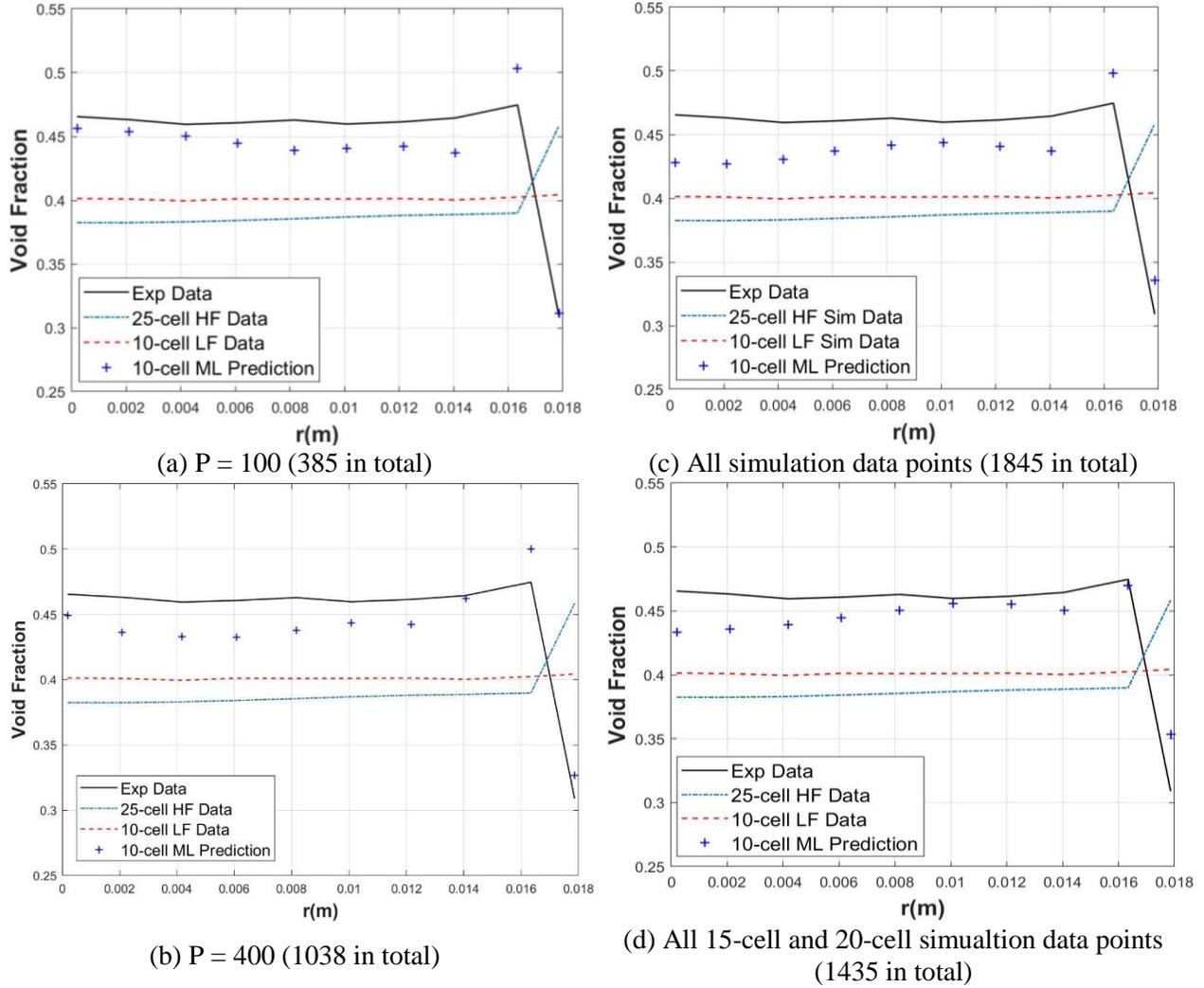

(a) P = 100 (385 in total)

(c) All simulation data points (1845 in total)

(b) P = 400 (1038 in total)

(d) All 15-cell and 20-cell simualtion data points (1435 in total)

Figure 10. Void fraction prediction of Test 2 for Case 7 with different datasets for DFNN training.

Figure 11 provides the data distribution of these target-similar data points in the other 41 cases with different injection velocities and void fractions when $P = 100$ and $P = 400$. The target case, Case 7, has the highest vapor-injection rate and void fraction and the lowest liquid-injection rate. Most of the data points come from Case 3–6, Case 11–14, and Case 20–21, which have injection velocities and void fractions similar to the local data of Case 7. However, all 41 cases are involved in the training-data selection. Even though global conditions (i.e., $u_g$, $u_l$ and $\alpha$) of some cases (e.g., Case 36, where three points are selected when $P = 400$) are quite different from those of Case 7, they still have some local data points which are more similar than some of globally similar cases (e.g., Case 6, where 5 points are not selected when $P = 400$). It denotes that even globally some cases are not similar to the target case, but some locally similar data points of these cases can still be used to inform the prediction of the target case. Instead of choosing training data based on global similarity, local similarity is used as the metric for the selection of "optimal" training data. In this way, all existing data can be sufficiently utilized for specific targets.

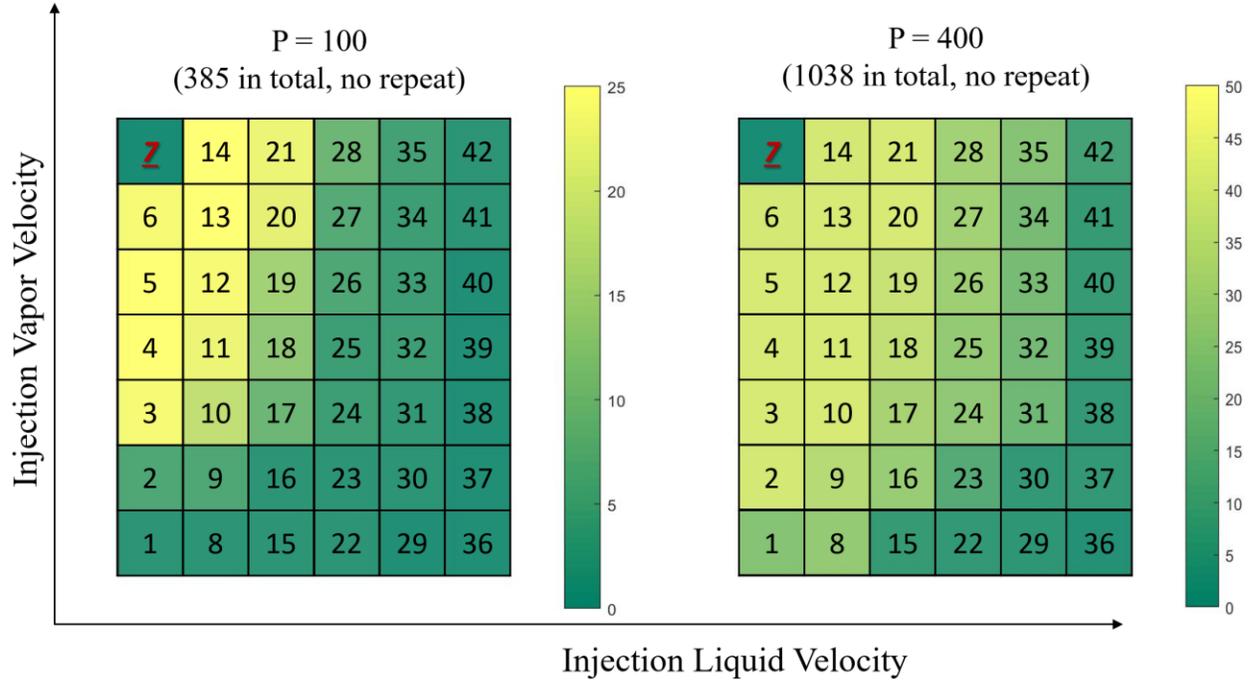

Figure 11. Distribution of target-similar data points in cases with different injection velocities.

### 4.3. Discussion on FSM predictive capability for extrapolation of flow regime

According to the flow-regime map for upward two-phase flow in vertical tubes developed by Mishima and Ishii (Mishima and Ishii, 1984), Cases 6 and 7 belong to slug flow while all other cases are classified as bubbly flow. Figure 12 compares these 42 cases with a two-phase bubbly-to-slug flow-regime map. Traditionally, different closure models for interfacial forces are developed and applied, respectively, for bubbly flow and slug flow. Considering BAMF model was developed for bubbly flow, its predictions for Case 6 and 7 do not well match respective experimental data, as shown in Figure 7 (d), (e) and (f), no matter whether using coarse or fine meshes. Depending on boundary conditions (superficial velocity), flow regime is still a global physical concept that describes the geometric distribution of liquid and vapor. Being limited by the small applicable range of closure models, each flow regime is analyzed separately and not consistent with others even though the boundary conditions are changing consistently (e.g., injection liquid and vapor velocities, void fractions). This implies that there is a scale gap among different global flow regimes; the closure models for interfacial forces do not have the predictive capability for the extrapolation of global flow regimes (i.e., scalability). By exploring local similarity among different global flow regimes, FSM is enabled with the scalability and consistency for different discontinuous flow regimes. Consistency indicates that even if LF predictions are made using the same simplified closure models and coarse meshes for different flow regimes, FSM has the capability correcting them to match HF predictions.

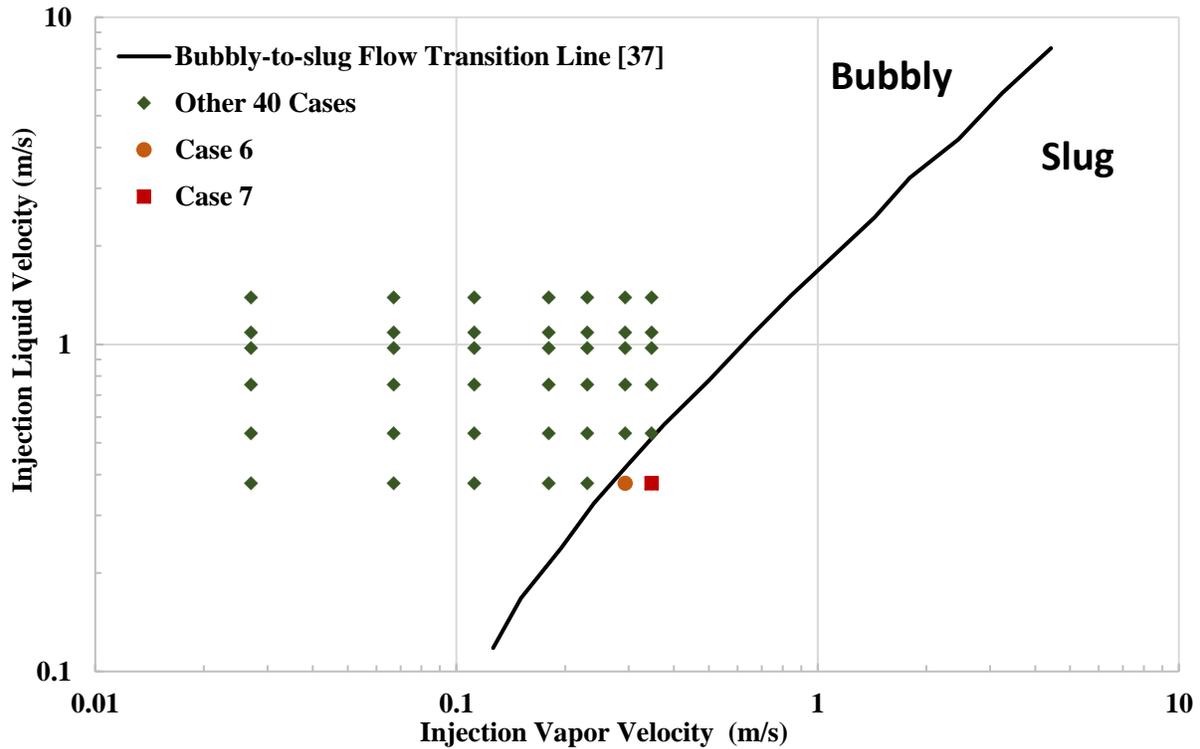

Figure 12. Locations of 42 cases in bubbly-to-slug flow regime map.

Figure 13 compares the LF predictions using LF-Type-I closures' respective experimental data and DFNN-corrected predictions using other 40 cases as training data for Cases 6 and 7. FSM utilizes DFNN as a surrogate modeling tool to capture the local patterns in the training data (40 cases in bubbly flow regime) and provides good predictions for the liquid-vapor behaviors in the target data (Cases 6 and 7 in slug-flow regime). The DFNN-corrected predictions for Case 7 are not as good as the ones in Figure 10 because data points in Case 6 are not included in the training dataset and only the data points in the other 40 bubbly flow cases are included.

Local similarity is enhanced by selecting the most similar data points for training ($P = 100$). The optimal training dataset for Cases 6 and 7, respectively, includes 386 and 382 data points in total. The distributions of target-similar data points in 40 training cases (two-phase bubbly flow) for Cases 6 and 7 (near two-phase bubbly-to-slug transition) are displayed in Figure 14. A major fraction of these data points comes from cases with similar injection superficial velocities, but every case in the bubbly flow regime contributes to the construction of training datasets.

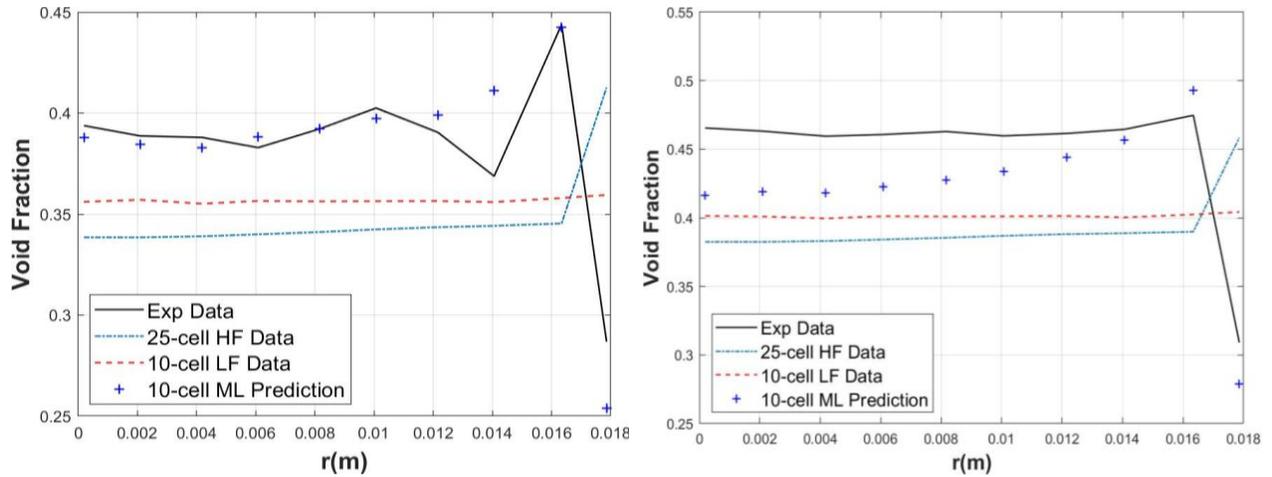

(a) Void fraction profile of Case 6　　　　　　　(b) Void fraction profile of Case 7

Figure 13. Comparisons of LF predictions using LF Type I closure, experimental data and DFNN-corrected data with the other 40 cases as training data for void-fraction profiles of Cases 6 and 7.

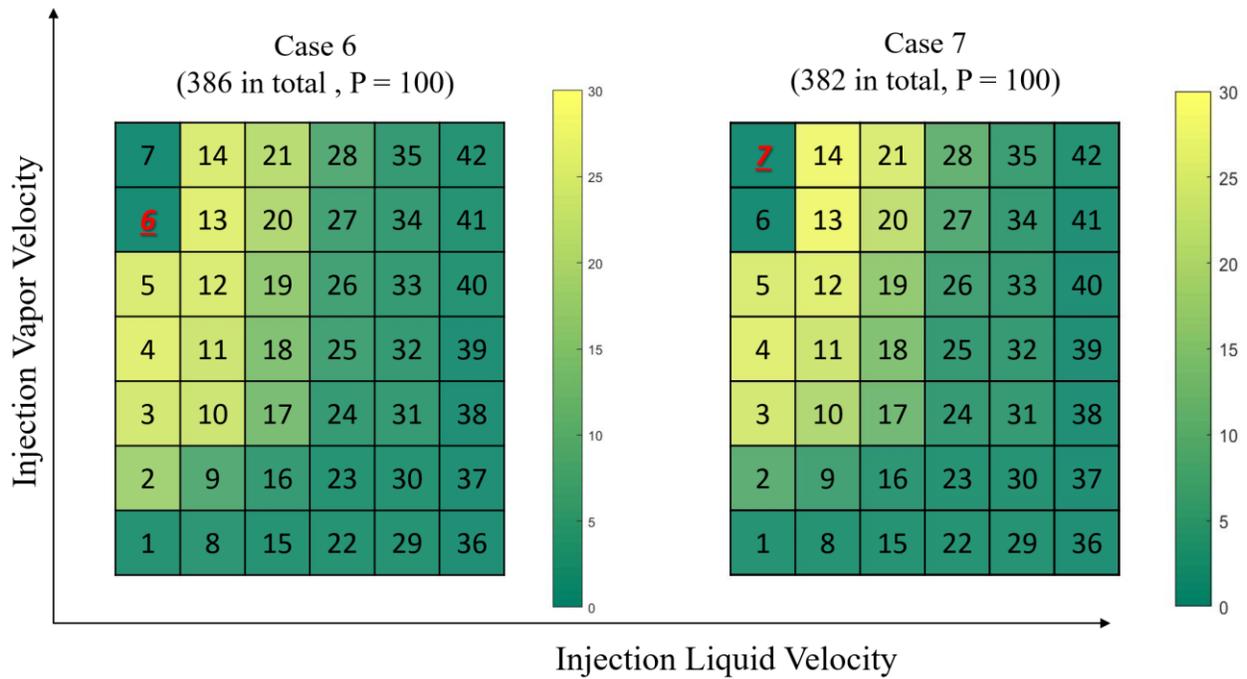

Figure 14. Distribution of target-similar data points in 40 training cases (two-phase bubbly flow) for Cases 6 and 7 (two-phase slug flow).

## 5. Conclusion

In this paper, FSM is applied to predict the simulation errors of two-phase flow using coarse-mesh CFD with simplified LF closure models. FSM takes model error, mesh-induced numerical error, and scaling distortions into consideration. Deep-learning technique is applied to explore the relationship between local physical-feature groups and simulation errors. The well-trained deep learning model can be considered as a surrogate of governing equations and closure models of coarse-mesh CFD.

As a data-driven approach for bridging the global scale gaps, FSM utilizes DFNN as a surrogate modeling tool to capture the local patterns of the training data (in two-phase bubbly flow regime) and predict the liquid-vapor behaviors of target data (near two-phase bubbly-to-slug transition). By exploring local similarity among different global conditions, FSM is enabled with the scalability and consistency. Scalability represents that FSM has the predictive capability for the extrapolation of mesh size, boundary condition, or flow regime, while consistency indicates that even if LF predictions are made using same simplified closure models and coarse meshes for different global conditions, FSM has the capability correcting them to match HF predictions. The scalability and consistency of FSM have been investigated in a two-phase pipe flow case study with different types of LF data and HF validation data. LF data is generated using different closure models and mesh sizes, HF data includes fine-mesh validated CFD predictions or experimental data. Results show that no matter which type of HF or LF data is used, FSM has the capability to capture their differences and correct LF data to match the selected HF validation data. However, the testing cases in the case study are near the two-phase bubbly-to-slug transition and close to bubbly flow regime. In future work, more studies will be performed to investigate the scalability of FSM by using more cases in slug flow regime for testing.

## Acknowledgment

This work is supported by the U.S. Department of Energy, under Department of Energy Idaho Operations Office Contract DE-AC07-05ID14517. Accordingly, the U.S. Government retains a nonexclusive, royalty-free license to publish or reproduce the published form of this contribution, or allow others to do so, for U.S. Government purposes. The authors also would like to acknowledge the support from Baglietto CFD Research Group at MIT.